\documentclass{PoS}

\title{Four hot DOGs eaten up with the EVN}

\ShortTitle{Hot DOGs with the EVN}

\author{\speaker{S\'andor Frey}\thanks{
The EVN is a joint facility of European, Chinese, South African, and other radio astronomy institutes funded by their national research councils. S.F. and K.\'E.G. thank the Hungarian Scientific Research Fund (OTKA NN110333) for support. The research leading to these results has received funding from the European Commission 7th Framework Programme (FP/2007-2013) under grant agreement No. 283393 (RadioNet3), and the China--Hungary Collaboration and Exchange Programme by the International Cooperation Bureau of the Chinese Academy of Sciences.}\\
        F\"OMI Satellite Geodetic Observatory, Hungary\\
        E-mail: \email{frey.sandor@fomi.hu}}

\author{Zsolt Paragi\\
        Joint Institute for VLBI in Europe, The Netherlands\\}

\author{Krisztina \'Eva Gab\'anyi\\
        F\"OMI Satellite Geodetic Observatory, Hungary\\
	Konkoly Observatory, MTA Research Centre for Astronomy and Earth Sciences, Hungary\\}

\author{Tao An\\
        Shanghai Astronomical Observatory, Chinese Academy of Sciences, China\\
        Key Laboratory of Radio Astronomy, Chinese Academy of Sciences, China}

\abstract{Hot dust-obscured galaxies (hot DOGs) are a rare class of hyperluminous infrared galaxies recently identified with the
{\em Wide-field Infrared Survey Explorer (WISE)} satellite. The majority of the $\sim$1000-member all-sky population should be at high redshifts ($z \sim 2-3$), at the peak of star formation in the history of the Universe. This class most likely represents a
short phase during galaxy merging and evolution, a transition from starburst- to AGN-dominated phases. For the first time, we observed four hot DOGs with known mJy-level radio emission using the European VLBI Network (EVN) at 1.7~GHz, in a hope to find compact radio features characteristic to AGN activity. All four target sources are detected at $\sim$15--30~mas angular resolution, confirming the presence of an active nucleus. The sources are spatially resolved, i.e. the flux density of the VLBI-detected components is smaller than the total flux density, suggesting that a fraction of the radio emission originates from larger-scale (partly starburst-related) activity. Here we show the preliminary results of our e-EVN observations made in 2014 February, and discuss WISE J1814+3412, an object with kpc-scale symmetric radio structure, in more detail.}

\FullConference{12th European VLBI Network Symposium and Users Meeting,\\
		7-10 October 2014\\
		Cagliari, Italy}

\begin{document}

\section{Introduction}

Among the primary targets of the {\em Wide-field Infrared Survey Explorer (WISE)} space telescope \cite{Wrig10} were the most luminous galaxies in the Universe. The WISE all-sky survey has been conducted at four infrared bands at wavelengths 3.4, 4.6, 12, and 22 $\mu$m (denoted by {\em W1}, {\em W2}, {\em W3}, and {\em W4}, respectively). In a search for hyperluminous infrared galaxies (HyLIRGs; $L_{\rm IR} > 10^{13} L_{\odot}$) from the WISE catalogue, the so-called {\em W1W2}-dropout objects became the best candidates \cite{Wu12,Eise12}. These are heavily obscured galaxies, very faint or even undetected in the {\em W1} and {\em W2} bands, but have significant flux density in {\em W3} and {\em W4}. At redshifts of $z \sim 2-3$ that correspond to the cosmological epoch of peak star formation, the near-infrared obscuration in the rest frame of the galaxies is expected to fall in the {\em W1} and {\em W2} bands, while the strong emission of the hot obscuring dust is detected at 12 and 22 $\mu$m. Indeed, spectroscopic observations of more than 100 {\em W1W2}-dropout galaxies \cite{Brid13,Eise12} revealed that the majority of these galaxies are at $z>1.6$, with the bulk of the sources at $z \sim 2-3$. The all-sky sample of {\em W1W2}-dropouts contains $\sim$1000 objects. Their surface density is at least 5 orders of magnitude smaller than that of {\em WISE} sources in general \cite{Eise12}, therefore these objects represent a rare type of hyperluminous galaxies, coined as hot DOGs (hot dust-obscured galaxies) \cite{Wu12}. Because of their rarity, hot DOGs likely signify a brief transitional phase of galaxy evolution. 

To reveal the nature of hot DOGs, there have been various studies of samples at different wavebands since their discovery a few years ago. The prototype of this class, the first such a source investigated in detail with optical, infrared, submillimetre and radio imaging, and optical spectroscopy was WISE J1814+3412 \cite{Eise12}. Its radio emission, significantly exceeding the far infrared--radio correlation, is considered as the strongest evidence for the presence of an active galactic nucleus (AGN) in the system. The observations of this Lyman break galaxy are well explained with a spectral energy distribution (SED) combined from an obscured AGN, a starburst, and an evolved stellar component. The star formation rate (SFR) is estimated to be $\sim$300~$M_{\odot}\,{\rm yr}^{-1}$ \cite{Eise12}. Submillimetre and infrared observations of a sample of hot DOGs indicate that the SEDs are dominated by hot (60--120~K) dust emission generated by powerful AGNs, but multiple components with different temperatures are also present \cite{Wu12}. Interferometric mm-wavelength observations with $\sim$1$^{\prime\prime}$ resolution led to the continuum detection of two unresolved sources out of the targeted three, and to the estimates of upper limits of molecular gas and cold dust contents in these galaxies \cite{Wu14}. Further submillimetre data of 10 hot DOGs confirmed that the SEDs cannot be well fitted by AGN templates alone, without considering extra dust extinction \cite{Jone14}. The first X-ray spectrum of a hot DOG (WISE J1835+4355) indicates the presence of a heavily obscured AGN \cite{Pico15}.   

The WISE-selected dusty Ly$\alpha$ emitters at high redshifts are typically radio-quiet \cite{Brid13}, but some of the known hot DOGs do show mJy-level radio emission at 1.4~GHz. This provides us with an intriguing possibility to directly confirm AGN-related radio emission in these sources since the dust in the galaxies is transparent at GHz frequencies. Moreover, the angular resolution achievable with the technique of very long baseline interferometry (VLBI) guarantees that the emission region is compact, and if a radio source is detected with VLBI from a sufficiently large distance (at redshifts well above $z$$\sim$0.1), then the corresponding luminosity clearly exceeds that of the brightest supernovae and/or supernove remnant complexes \cite{Alex12,Midd13}. Therefore VLBI offers a sharp and unobscured view of the compact AGN-related structures if they are indeed present in radio-emitting hot DOGs.

We selected four hot DOGs from the sample of \cite{Wu12} which are detected in the radio band. Three of them (WISE J0757+5113, WISE J1146+4129, WISE J1603+2745) are listed in the Very Large Array (VLA) Faint Images of the Radio Sky at Twenty-cm (FIRST) survey catalogue \cite{Whit97} as unresolved ($<5^{\prime\prime}$) objects, with integral flux densities of $\sim2-4$~mJy. The fourth hot DOG (WISE J1814+3412) was observed and detected as an unresolved source ($<9^{\prime\prime}$) with the Expanded VLA at two different frequencies, 4.5~GHz and 7.9~GHz \cite{Eise12}. From the measured spectral index, the extrapolated total 1.4-GHz flux density of WISE J1814+3412 is 1.4~mJy \cite{Eise12}.  

Here we present preliminary results of our VLBI study of these four hot DOGs at 1.7~GHz. The observations were performed with the European VLBI Network (EVN) on 2014 February 21-22. Throughout this paper, we assume a flat cosmological model with $H_{\rm 0}$=70~km~s$^{-1}$~Mpc$^{-1}$, $\Omega_{\rm m}$=0.3, and $\Omega_{\Lambda}=$0.7.

\section{Observations and results}

A network of eight radio telescopes (Effelsberg in Germany, the Jodrell Bank Lovell Telescope in the UK, Medicina and Noto in Italy, Onsala in Sweden, Toru\'n in Poland, the Westerbork Synthesis Radio Telescope in the Netherlands, and Sheshan in China) participated in the EVN observations. The experiment EF025 lasted for 14\,h and was performed in e-VLBI mode. The data were streamed from the radio telescopes to the central data processor \cite{Keim15} in real time. The four weak target sources were observed in phase-reference mode, using nearby bright, compact phase calibrator sources located within $\sim$3$^{\circ}$ in the sky. Further details of the observations, data calibration and imaging will be given elsewhere (S. Frey et al. 2015, in preparation).

\begin{figure}
\centering
  \includegraphics[bb=70 125 524 671, width=90mm, angle=270, clip=]{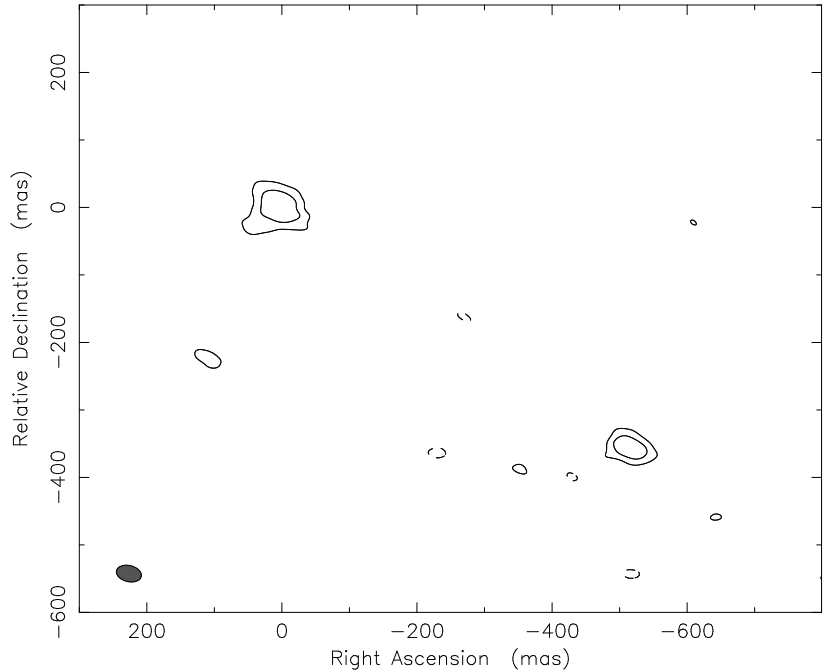}
  \caption{
The naturally-weighted 1.7-GHz EVN image of WISE J1814+3412. The peak brightness is 157~$\mu$Jy~beam$^{-1}$, the contours are drawn at $\pm$47 and +94~$\mu$Jy~beam$^{-1}$. The lowest contour is at $\sim$4$\sigma$ image noise level. The Gaussian restoring beam (full width at half maximum 37.6\,mas $\times$ 23.9\,mas, major axis position angle 76$^{\circ}$) is indicated with a filled ellipse in the lower-left corner. We show the coordinates relative to the brightness peak. The absolute position of the north-eastern component is 18$^{\rm h}$\,14$^{\rm m}$\,17.3077$^{\rm s}$ and declination 34$^{\circ}$\,12$^{\prime}$\,25.438$^{\prime\prime}$, and is accurate to $\sim$2~mas in both coordinates.}
  \label{images}
\end{figure}

The four hot DOGs were all detected with the EVN but heavily resolved on the longest intercontinental baselines between Europe and China. As an example, we show here the 1.7-GHz VLBI image of WISE J1814+3412 in Fig.~\ref{images}, where the data from Sheshan antenna were omitted and natural weighting was applied. On the other hand, the four hot DOGs were completely resolved out (remained undetected) in the uniformly-weighted dirty images. These would have provided the highest angular resolution, about 2\,mas~$\times$~9\,mas with the full array in our experiment. The non-detection implies about 1\,mJy\,beam$^{-1}$ ($\sim$5$\sigma$) brightness upper limit, and allows us to estimate the maximum brightness temperature, $\sim$$10^{8}$\,K, of any putative mas-scale compact radio component in these galaxies. This clearly excludes any compact radio jet emission that is relativistically beamed towards the observer from these hot DOGs. The flux densities obtained with VLBI on angular scales of $\sim$10\,mas are smaller than the total flux densities measured at different epochs for these sources. While source variability is not excluded, this is probably due to resolving out large-scale radio structure on $>$100~mas scales. The lack of variability would be consistent with the absence of compact Doppler-boosted AGN jets and the resolved radio structures seen in the EVN images. The typical rest-frame 1.4-GHz radio powers of the detected VLBI components are in the order of $10^{25}$\,W\,Hz$^{-1}$. 

The 1.7-GHz VLBI image of the hot DOG ``prototype'' WISE J1814+3412 \cite{Eise12} shows a symmetric structure, with an angular separation of $\sim$630 milli-arcseconds (mas) at a position angle $\sim$56$^{\circ}$ (Fig.~\ref{images}), corresponding to 5.1~kpc projected linear separation at $z=2.452$. Based on the morphology, the component separation and the radio powers ($\sim$$10^{25}$\,W\,Hz$^{-1}$), the radio structure of WISE J1814+3412 can be interpreted as two lobes of a medium-sized symmetric object (MSO). Low-power MSOs represent an early phase of radio-loud AGN development, possibly becoming FR-I type radio sources at a later stage \cite{An12}. The MSO interpretation is also consitent with the steep overall radio spectrum of the source (spectral index $\alpha=-0.8$, defined as $S\propto\nu^{\alpha}$, where $S$ is the flux density and $\nu$ the frequency) \cite{Eise12}. Interestingly, the line connecting the radio source pair is nearly perpendicular to the characteristic position angle of the Ly$\alpha$-emitting region ($\sim$140$^{\circ}$) that extends to more than 30 kpc in the galaxy \cite{Eise12}. 

An alternative explanation of the radio structure in WISE J1814+3412 would be a dual AGN (see e.g. \cite{Frey12} and references therein). During their formation and growth, galaxies as well as their nuclei naturally go through mergers. If the accretion onto the central supermassive black holes of two merging galaxies is maintained at the same time, dual AGN systems may be observed. However, it is difficult to find unequivocal observational evidence for kpc-scale dual AGN. In the case of this object, there is no other indication of duality, although WISE J1814+3412 is accompanied by a quasar (5.2$^{\prime\prime}$ separation; within 42~kpc projected linear distance) and another Lyman break galaxy (3.8$^{\prime\prime}$ separation) at the same redshift \cite{Eise12}. It is also suggested that there is an overdensity of serendipitous sources around hot DOGs on arcmin scale \cite{Jone14}. We note that \cite{Eise12} found a positional shift between the optical ($g^{\prime}$) image centroid and the location of WISE J1814+3412 in the {\em Spitzer} $K_S$ (2.15~$\mu$m) infrared image, with the optical position being displaced eastward by about 0.5$^{\prime\prime}$. They interpret this as a result of extinction. However, the amount of this shift (but not exactly the direction) is comparable to the separation of the two radio components in Fig.~\ref{images}.   

From our VLBI data, neglecting any flux density variability, we find that $\sim80$\% of the total 1.4-GHz radio flux density of WISE J1814+3412 \cite{Eise12} is contained in the two compact components in Fig.~\ref{images}. The remaining $\sim0.3$~mJy (corresponding to $1.1 \times 10^{25}$\,W\,Hz$^{-1}$ power) is resolved out by the interferometer and is explained by emission extended to larger scales -- either startburst-related or diffuse AGN lobe emission. Considering amplitude calibration uncertainties, this value could be in error by up to 50\%. 
Nevertheless, a rough estimation of the upper limit to the SFR is possible if we assume that all radio emission resolved out by the EVN is associated with star formation only. Applying the relation between the SFR and the radio power \cite{Hopk03}, we obtain SFR$<$6000~$M_{\odot}\,{\rm yr}^{-1}$. This upper limit is an order of magnitude higher than the maximum SFR estimated by \cite{Eise12}. It is therefore most likely that the majority of the ``missing'' VLBI flux density can be attributed to diffuse AGN emission in WISE J1814+3412.

\section{Summary}

Radio emission is free from dust obscuration, and VLBI is a unique tool to study radio AGN at very high resolution. Several pieces of evidence indicate that hot DOGs -- a rare type of hyperluminous infrared galaxies at high redshifts -- are powered by hidden AGNs whose radiation heats the dust in their host galaxies. The majority of hot DOGs are not radio-loud, but some of them show mJy-level flux densities at GHz frequencies. With new 1.7-GHz VLBI imaging observations of four hot DOGs selected by their radio emission, we found that they indeed contain AGN, supporting the general picture of hot DOGs. The source WISE J1814+3412 which has been extensively studied at different wavebands shows a double radio structure in our VLBI image. The nearly symmetrical components separated by $\sim$630~mas (5.1~kpc) may be the lobes of a young, expanding radio source, an MSO. Alternatively, the components might be interpreted as a dual AGN. Future sensitive VLBI observations of a larger sample of radio-emitting hot DOGs could provide valuable information to better understand this type of objects.

\end{document}